\documentclass{optica-article}
\usepackage{makecell}
\usepackage{pifont}
\usepackage{graphicx}
\usepackage{float}
\journal{opticajournal} 

\articletype{Research Article}

\begin{document}

\title{Narrow-linewidth 852-nm DBR-LD with self-injection lock based on high-finesse optical cavity filtering}

\author{ Lili Hao $^{1}$, Rui Chang $^{1}$, Xiaokai Hou $^{1}$, Jun He $^{1,2}$ and Junmin Wang $^{1,2,*}$}

\address{\authormark{1}State Key Laboratory of Quantum Optics and Quantum Optics Devices, and Institute of Opto-Electronics, Shanxi University, Tai Yuan 030006, Shanxi Province, China\\
\authormark{2}Collaborative Innovation Center of Extreme Optics, Shanxi University, Tai Yuan 030006, Shanxi Province, China\\}

\email{\authormark{*}wwjjmm@sxu.edu.cn} 


\begin{abstract*} 
Narrow-linewidth lasers have a high spectral purity, long coherent length, and low phase noise, so they have important applications in atomic clocks, precision measurement, and quantum computing. We inject a transmitted laser from a narrow-linewidth ($\sim15$ kHz) flat-concave Fabry--Perot (F-P) cavity made from ultra-low expansion (ULE) optical glass into an 852 nm distributed Bragg reflector-type laser diode (DBR-LD), of which the comprehensive linewidth is 1.67 MHz for the free running case. With an increase in the self-injection power, the laser linewidth gradually narrows, and the injection locking current range gradually increases. The narrowest linewidth measured by the delayed frequency-shifted self-heterodyne (DFSSH) method is about 365 Hz, which is about $1/4500$ of the linewidth for the free running case. Moreover, to characterize the laser phase noise, we use a detuned F-P cavity to measure the conversion signal from the laser phase noise to the intensity noise for both the free running case and the self-injection lock case. The laser phase noise for the self-injection lock case is significantly suppressed in the analysis frequency range of 0.1--10 MHz compared to the free running case. In particular, the phase noise is suppressed by more than 30 dB at an analysis frequency of 100 kHz.

\end{abstract*}
\section{Introduction}

Laser diodes have the characteristics of low power consumption, high efficiency, easy integration, and a wide range of continuous tuning, so they have important applications in atomic physics \cite{ref-1}, quantum optics \cite{ref-2}, and laser spectroscopy \cite{ref-3}. With the development of science and technology, the narrow linewidth, high stability, and low-noise laser have become necessary conditions for atomic clocks \cite{ref-4}, precision measurement \cite{ref-5}, and quantum computing \cite{ref-6}. 

After the first laser diode was successfully developed, in 1964, Crowe and Craig \cite{ref-7} added a tunable external cavity to the laser diode. In 1980, Lang and Kobayashi \cite{ref-8} studied the influence of laser external cavity feedback on the performance of the laser diode. The external cavity effectively narrowed the laser linewidth and expanded the laser tunable range. They also demonstrated the effects of the external cavity length and injection current on the output laser of the external cavity laser diode. In 1981, Fleming and Mooradian \cite{ref-9} first used diffraction gratings as an external cavity, and the laser linewidth was narrowed to 1.5 MHz. With the development of technology, the current external cavity laser diode with a laser linewidth of 100 kHz could also no longer meet the needs of current experiments in terms of the Rydberg excitation, atomic clock, etc. Therefore, the electrical feedback method \cite{ref-10,ref-11} or optical feedback method \cite{ref-12,ref-13} was proposed to effectively narrow the laser linewidth and reduce the phase noise. The common way to narrow the linewidth by electrical feedback is to use a high-finesse optical cavity as the frequency reference, and the laser diode is locked on the cavity using Pound–Drever–Hall (PDH) \cite{ref-14,ref-15} technology \cite{ref-16,ref-17}. The laser linewidth can be narrowed to the Hz \cite{ref-18} and even to mHz level \cite{ref-19}, but its circuit and optical path setup requirements are very strict, and due to its complex circuit, the introduction of other electrical noise is inevitable. Compared with the electrical feedback method, optical feedback technology \cite{ref-20} uses the inherent high-frequency sensitivity of the laser diode to the injected laser, which can effectively narrow the laser linewidth and reduce the phase noise, making it an ideal way to obtain high-quality laser sources.

In order to obtain a narrower linewidth, the injected laser should have a narrower linewidth and a lower phase noise, so the high-finesse F-P cavity is widely used as a good filtering element. {For example, in 2012, the Zang Erjun group \cite{ref-21} used a high-finesse optical cavity to achieve the self-injection locking of a 689 nm external cavity laser diode, and the laser linewidth was narrowed to 100 Hz, while the laser phase noise was reduced by more than 50 dB in the frequency range of 10 Hz to 10 kHz. In 2023, Liang Wei group \cite{ref-22} realized the miniaturization of the self-injection locking experimental device using a high-finesse optical cavity. The laser linewidth narrowed to 80 Hz. In this paper, we use self-injection locking technology to obtain an 852 nm distributed Bragg reflector-type laser diode (DBR-LD) with a narrow linewidth and low phase noise. We select a suitable lens to match the high-finesse F-P cavity and tune the laser frequency to the cavity resonance frequency. We use the transmitted laser to carry out the self-injection lock, and the measured DBR-LD laser linewidth is about 365 Hz, which is about $1/4500$ of the linewidth for the free running case. Then, we test the phase noise of the laser for the free running case and the self-injection lock case for the 0.1--10 MHz analysis frequency range using a detuned F-P cavity method. At the 100 kHz analysis frequency, the laser phase noise is significantly reduced by 30 dB.
\section{Experimental principle}
\subsection{Self-injection lock of the laser diode}

Optical injection is the process of injecting a laser with a certain mode into the laser cavity, which the injected laser competes with the free running longitudinal mode inside the laser to consume carriers and affect its output characteristics \cite{ref-23,ref-24}. When the intensity of the injected laser field increases, the number of carriers consumed by the corresponding mode gradually increases. When the number of carriers is lower than the threshold, the internal free running mode will not be excited, the laser output's laser frequency will be the same as that of the injected laser, and the phase difference will be constant. At this time, the injection lock is realized. The following uses the single-mode field model based on the laser diode rate equation to analyze the injection locking principle.

Without injection, the laser is in free running, and its mode is expressed as~\cite{ref-25,ref-26}:
\begin{equation}
\label{equ1}
E_{s}=\sqrt{P_{s}(t)} e^{i(\omega_st+\Phi_s(t))}   
\end{equation}

The longitudinal mode of the injected laser is expressed as:
\begin{equation}
\label{equ2}
E_{i}=\sqrt{P_{i}(t)} e^{i(\omega_it+\Phi_i(t))}   
\end{equation}
where subscript $s$ is the laser longitudinal mode for the free running case, and the subscript $i$ is the injection laser longitudinal mode, $P$ stands for power, $\omega$ for frequency, and $\Phi$ \mbox{for phase.}

Suppose that the injection laser mode at this time corresponds to one of the many cavity modes of the laser:
\begin{equation}
\label{equ3}
E_{ss}=\sqrt{P_{ss}(t)} e^{i(\omega_{ss}t+\Phi_{ss}(t))}   
\end{equation}

Suppose that $G_s$ and $G_{ss}$ are gain factors for the $E_s(t)$ and $E_{ss}(t)$ modes, respectively, $\delta_s$ and $\delta_{ss}$ are the loss factors of the two modes of $E_s(t)$ and $E_{ss}(t)$, respectively. The subscript $ss$ stands for the cavity mode of the laser corresponding to the injected laser longitudinal mode. $\alpha$ is a linewidth gain factor that depends on carrier-induced changes in the refractive index. The same cavity length inside the laser can be compatible with many longitudinal modes, and only the longitudinal mode of its free running mode and the side mode consistent with the frequency of the injected laser are considered:
\begin{equation}
\label{equ4}
\frac{dE_s(t)}{dt}=\frac{1}{2} (1-i\alpha)(G_s-\delta_s)E_s(t)   
\end{equation}
\begin{equation}
\label{equ5}
\frac{dE_{ss}(t)}{dt}=\frac{1}{2} (1-i\alpha)(G_{ss}-\delta_{ss})E_{ss}(t)   
\end{equation}

The carrier rate equation within a laser diode is expressed as:
\begin{equation}
\label{equ6}
\frac{dN(t)}{dt}=J-\frac{N(t)}{\tau_s}-G_{s}E_{s}(t)-G_{ss}E_{ss}(t)  
\end{equation}
where $N(t)$ is the number of carriers, $\tau_s$ is the carrier's lifetime, and $J$ is the injection current.

When injected, the injected laser $E_{ss}$ interacts only with the side mode, and the photons of these two modes compete by consuming carriers, saturating the gain factor. To simplify the theory, we suppose $G_s$=$G_{ss}$=$G$. Substituting equations (1), (2), and (3) into (4), (5), and (6), respectively, yields the following results:
\begin{equation}
\label{equ7}
\frac{dP_s(t)}{dt}=(G-\delta_s)P_s(t)  
\end{equation}
\begin{equation}
\label{equ8}
\frac{d\Phi_s(t)}{dt}=\frac{\alpha}{2}(G-\delta_s)  
\end{equation}
\begin{equation}
\label{equ9}
\frac{dP_{ss}(t)}{dt}=(G-\delta_{ss})P_s(t)+2\gamma_c\sqrt{P_i(t)P_s(t)}cos\Phi_{ss}(t)
\end{equation}
\begin{equation}
\label{equ10}
\frac{d\Phi_{ss}(t)}{dt}=\Delta{\omega}+\frac{\alpha}{2}(G-\delta_{ss})-\gamma_c\sqrt{\frac{P_i(t)}{P_s(t)}}sin\Phi_{ss}(t) 
\end{equation}
\begin{equation}
\label{equ11}
\frac{dN(t)}{dt}=J-\frac{N(t)}{\tau_s}-G[P_s(t)+P_{ss}(t)+2\sqrt{P_s(t)P_{ss}(t)cos(\Delta{\omega}+\Phi)}]
\end{equation}
where $\Delta{\omega}$ = $\omega_i$ $-$ $\omega_s$ is the frequency difference between two lasers, $\Phi$ = $\Phi_i$ $-$ $\Phi_{ss}$ the phase difference, which determines the phase following effect of the injection lock, and $\gamma_c$ is the injection coupling parameter.

When the laser reaches a stable operating state, $\frac{dP_{ss}(t)}{dt}=0$, $\frac{d\Phi_{ss}(t)}{dt}=0$.
\begin{equation}
\label{equ12}
-\gamma_c\sqrt{\frac{P_i}{P_s}}\le\Delta{\omega}\le\gamma_c\sqrt{\frac{P_i}{P_s}}\sqrt{1+\alpha^2}
\end{equation}

The locking range is proportional to the square root of the injection power and is related to the linewidth gain factor $\alpha$ and the injection coupling parameter $\gamma_c$.

Next, we consider how the optical feedback changes the laser linewidth. The laser linewidth is caused by random phase fluctuations caused by spontaneous radiation. When there is an external feedback laser and stable operation, the laser linewidth mainly depends on the external feedback laser and its relative phase, resulting in the laser linewidth of a longitudinal mode that is significantly wider or narrower. The effect of optical feedback on the laser linewidth can be expressed as the following equation \cite{ref-27}:
\begin{equation}
\label{equ13}
\Delta{f}=\frac{\Delta{f_0}}{[1+Ccos(\Phi_i+tan^{-1}\alpha)]}
\end{equation}
where $\Delta{f_o}$ is the laser linewidth for the free running case, $\Delta{f}$ is the laser linewidth when there is optical feedback, $C$ is the feedback parameter, and if 
$\Phi_i + tan^{-1}\alpha =2m \pi$, the laser linewidth decreases over time $(1+C)^2$. Therefore, the linewidth of the output laser can be changed by injecting laser on the phase.
\subsection{The Delayed Frequency-Shifted Self-Heterodyne (DFSSH) Method for Laser Linewidth~Measurement}

The basic principle of measuring the laser linewidth by the delayed frequency-shifted self-heterodyne (DFSSH) method \cite{ref-28,ref-29} is an unequal arm interferometer used to generate two beams with different delay times for the heterodyne. The DFSSH beatnote signal carries the laser frequency noise, and the laser linewidth can be obtained by analyzing the signal. In order to avoid the disturbance of low-frequency noise, the acousto-optic modulator (AOM) is used to shift the frequency in the interferometer, and the signal is detected by the detector, and its power spectra are recorded and analyzed by an RF spectrum analyzer \cite{ref-30}:
\begin{equation}
\label{equ14}
S(f,\Delta{f)}=S_1S_2+S_3
\end{equation}
\begin{equation}
\label{equ15}
S_1=\frac{P_0^2}{4\pi}\frac{\Delta{f}}{\Delta{f^2}+(f-\Omega)^2}
\end{equation}
\begin{equation}
\label{equ16}
S_2=1-exp(-2\pi\Delta{f}\tau_d)[cos[2\pi(f-{\Omega)\tau_d}]+\Delta{f}\frac{sin[2\pi(f-{\Omega)\tau_d}]}{f-{\Omega}}]
\end{equation}
\begin{equation}
\label{equ17}
s_3=\frac{\pi{P_0^2}}{2}exp(-2\pi{\Delta{f}\tau_d})\delta({f-{\Omega}})
\end{equation}

Among them, $f$ is the measurement frequency, $\Delta{f}$ is the laser linewidth, $\tau_d$ is the fiber delay time, and $\Omega$ is the AOM frequency shift. $S_1$ is the Lorentz function, and its spectral line shape is not affected by the fiber length. $S_3$ is a Delta function pulse peak, and its spectral lines form a narrow spike at the center frequency. The $S_2$ function can reflect the difference in the spectral line envelope, and the oscillation of the two wings of the power spectrum is caused by the e-index part of the $S_2$ function. $S_2$ is a periodic function. Its amplitude and period are affected by the length of the delay fiber, where the longer the length of the delay fiber, the smaller the amplitude and period of the function and the closer the first minimum point is to the center frequency. When the length of the delay fiber increases to a certain extent, it is generally believed that when the fiber delay time is greater than six times the laser coherent time, the oscillation of the two wings of the power spectrum disappears. At this time, when the output power spectrum of the laser is Lorentzian, its beatnote signal is still Lorentzian, and the half-maximum full width of the beatnote signal is 2 times the laser linewidth. When the output power spectrum of the laser is Gaussian, the beatnote signal is still Gaussian, and the half-maximum full width of the beatnote signal is $\sqrt{2}$ times the laser linewidth. 
\subsection{The detuned F-P cavity method for laser phase noise measurement}

Phase noise is one of the important indicators used to evaluate the laser diodes. And, low-phase noise lasers are widely used in experiments on the interaction between the laser and atoms. However, because phase noise cannot be measured directly, people have proposed the measurement of laser phase noise indirectly through various conversion methods. The detuned F-P cavity method can usually be used to convert laser phase noise into intensity noise for measurement \cite{ref-31}. We assume that a beam is injected into the F-P cavity, and the reflection coefficient and transmission coefficient of the incident mirror to the injected laser are $r_1$ and $t_1$, respectively. Ideally, they should satisfy $r_1^2+t_1^2=1$. And, the reflection coefficient of the output mirror to the injected laser is $r_2$. Suppose $E_{in}$ and $E_{out}$ are the input and output fields, respectively, $E_{cav}$ and $E'_{cav}$ are the fields within the cavity at the incident and outgoing exit, respectively. From the input--output relationship, we can obtain the following equation \cite{ref-32}:
\begin{equation}
\label{equ18}
{{t}_{1}}{{E}_{in}}={{E}_{cav}}-{{r}_{1}}E_{cav}^{'}
\end{equation}
\begin{equation}
\label{equ19}
{{E}_{out}}={{t}_{1}}E_{cav}^{'}-{{r}_{1}}{{E}_{in}}
\end{equation}
\begin{equation}
\label{equ20}
E_{cav}^{'}={{r}_{2}}{{E}_{cav}}\exp (-i\Phi )
\end{equation}
where $\Phi$ is the phase delay of laser going back and forth in the cavity. The variables $p$ and $q$ are introduced, which are proportional to the intensity and phase of the laser field, respectively.
\begin{equation}
\label{equ21}
p=\delta (\omega )+{{\delta }^{*}}(-\omega)
\end{equation}
\begin{equation}
\label{equ22}
q=-i[\delta (\omega )-{{\delta }^{*}}(-\omega)]
\end{equation}

The power spectrum of the reflected field of the F-P cavity can be calculated as:
\begin{equation}
\label{equ23}
S_{1}^{out}=S_{i}^{out}+S_{d}^{out}
\end{equation}
\begin{equation}
\label{equ24}
S_{i}^{out}={{\left| \frac{1}{2}{M}\{A+B\} \right|}^{2}}{{\left| {{p}_{in}} \right|}^{2}}
\end{equation}
\begin{equation}
\label{equ25}
S_{d}^{out}={{\left| \frac{1}{2}{M}\{A-B\} \right|}^{2}}{{\left| {{q}_{in}} \right|}^{2}}
\end{equation}
\begin{equation}
\label{equ26}
M=\overline{\frac{1+r_{1}^{2}r_{2}^{2}-2{{r}_{1}}{{r}_{2}}\cos \Phi }{r_{1}^{2}+r_{2}^{2}-2{{r}_{1}}{{r}_{2}}\cos \Phi }}
\end{equation}
\begin{equation}
\label{equ27}
A=\frac{{{r}_{2}}\exp (i\Phi )-{{r}_{1}}}{1-{{r}_{1}}{{r}_{2}}\exp (i\Phi )}\frac{{{r}_{2}}\exp [-i(\Phi -\Omega )]-{{r}_{1}}}{1-{{r}_{1}}{{r}_{2}}\exp [-i(\Phi -\Omega )]}
\end{equation}
\begin{equation}
\label{equ28}
B=\frac{{{r}_{2}}\exp (-i\Phi )-{{r}_{1}}}{1-{{r}_{1}}{{r}_{2}}\exp (-i\Phi )}\frac{{{r}_{2}}\exp [i(\Phi -\Omega )]-{{r}_{1}}}{1-{{r}_{1}}{{r}_{2}}\exp [i(\Phi -\Omega )]}
\end{equation}

Among them, $S_{i}^{out}$ is the intensity noise component of the laser field, and $S_{d}^{out}$ is the phase noise component of the laser field. $M$, $A$, and $B$ are auxiliary parameters, which have no practical significance.

It can be seen from Equation (23) that the intensity noise of the output laser contains both the intensity noise and the phase noise for the input laser through the cavity conversion device. When the F-P cavity frequency resonates with the input laser frequency, the noise converted by the intensity fluctuation is the largest, and the noise converted by its phase fluctuation is the smallest. When the cavity frequency slightly detunes with the input laser frequency, the noise converted by the intensity fluctuation decreases, and the noise converted by the phase fluctuation increases, until the noise converted by the intensity fluctuation is minimized, and the noise converted by phase fluctuation is maximum. When the detuning of the F-P cavity increases further, the noise converted by the intensity fluctuation is increased, and the noise converted by the phase fluctuation is reduced, until the frequency of the laser is far detuned of the resonant frequency of the F-P cavity. The noise of the reflected laser is mainly the intensity fluctuation. If the length of the F-P cavity is scanned periodically, the noise characteristics of the reflected laser of the cavity converted by these two types of noise also change periodically; and so, the phase--amplitude conversion is achieved. The conversion curve of the laser intensity noise to the phase noise (M-type conversion curve) can be observed on the spectrum analyzer.
\section{Experimental setup}

We perform a self-injection lock experiment of 852 nm DBR-LD using a high-finesse F-P cavity placed in an ultra-high vacuum as the filter cavity. Figure 1 shows the experimental setup for the DBR-LD self-injection lock and the DFSSH method used for the laser linewidth measurement. The DBR-LD used in our experiment was purchased from Photodigm (PH852DBR120T8), the material was an AlGaAs semiconductor, the package was TO-8 packaged, and its maximum output power was about 120 mW. The DBR-LD output laser becomes an approximate circular spot after passing through the half-wave plate ($\lambda$/2) and an anamorphic prism pair (APP). Then, the output laser beam passes through a specially designed optical isolator composed of a high-extinction ratio Glan--Taylor prism (GT), a Faraday rotator (FR), $\lambda$/2 and a polarization beam splitter cube (PBS). It allows the self-injected S-polarized laser beam to be fed back to the laser diode, and the P-polarized laser beam can be isolated. In order to further avoid the influence of the P-polarized laser beam, an optical isolator of -60 dB is used. Then, the laser beam is divided into two parts by $\lambda$/2 and PBS. One beam is used for the self-injection lock, which is shown in Figure 1(a). We match the high-finesse F-P cavity with two high-reflectivity mirrors and a plano-convex lens. The transmitted laser is divided into two beams after passing through $\lambda$/2 and PBS. One beam is used to monitor the cavity mode for the free running case and the self-injection locking case, and the other beam is used for the self-injection lock into the DBR-LD. In order to avoid other problems caused by the S-polarized laser beam, we built a simple isolator composed of PBS and a quarter-wave plate ($\lambda$/4) in the self-injection path. The other beam is used to measure the laser linewidth by the DFSSH method, as shown in Figure 1(b). 

\begin{figure}[H]
\centering
\includegraphics[width=7cm,height=5.5cm]{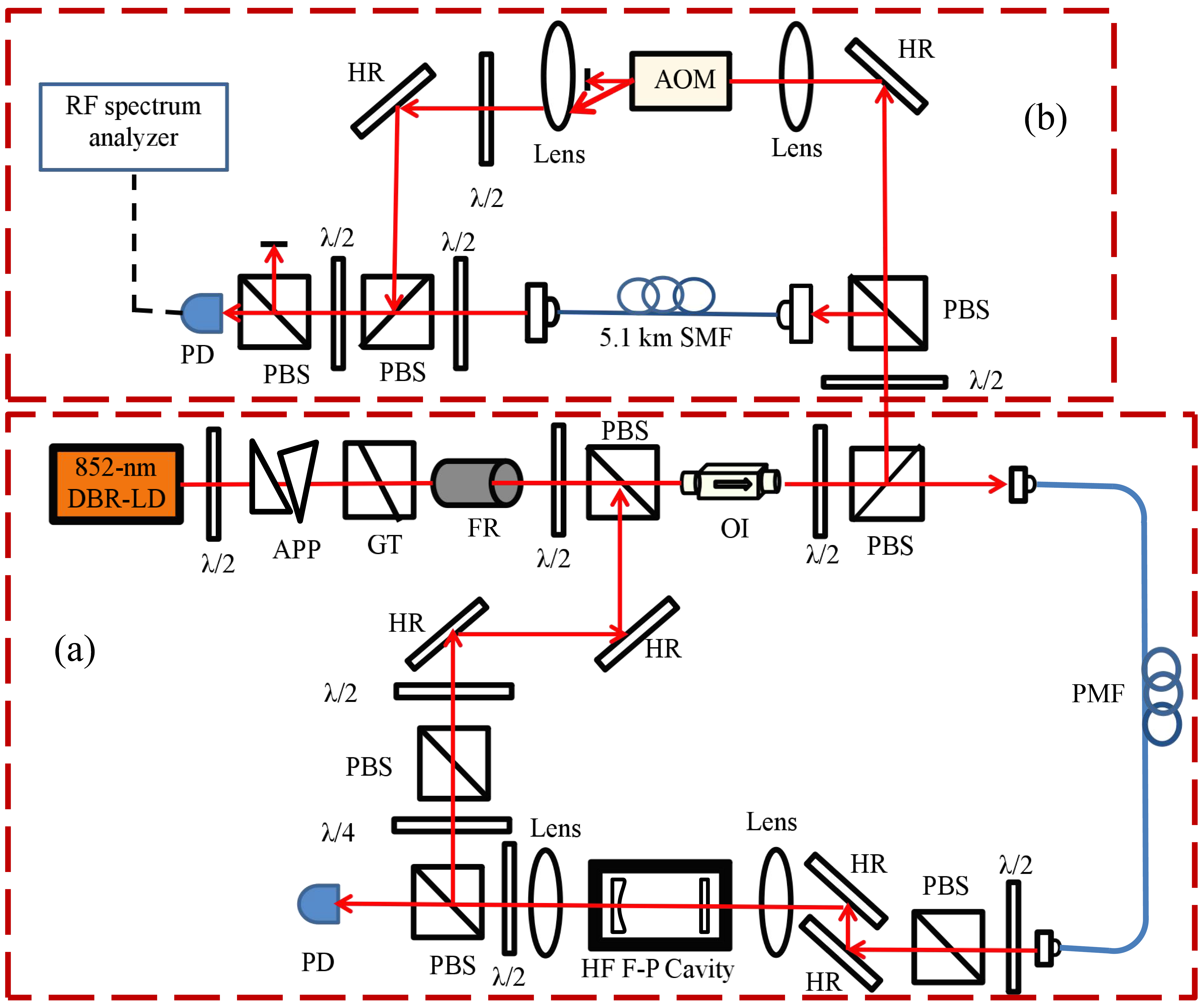}
\caption{Experimental setup diagram. (\textbf{a}) DBR-LD self-injection lock. (\textbf{b}) DFSSH method for linewidth measurement. $\lambda$/2: half-wave plate; APP: anamorphic prism pair; GT: Glan-Taylor prism; FR: Faraday rotator; OI: optical isolator; PBS: polarization beam splitter cube; PMF: polarization-maintaining fiber; HF F-P cavity: high-finesse Fabry-Perot cavity; $\lambda$/4: quarter-wave plate; HR: high-reflectivity mirror; SMF: single-mode fiber; AOM: acousto-optic modulator; PD: photodetector.}
\label{f1}
\end{figure}
\section{Experimental Results}
\subsection{DBR-LD Laser Linewidth of the Free Running and Self-Injection Lock Case}

Figure \ref{f2}a shows the DFSSH beatnote signal of the 852 nm DBR-LD laser for the free running case. The laser beam is divided into two beams: one is shifted by 80 MHz by the AOM, and the other is delayed by the 5.1 km single-mode fiber (SMF). The two beams are combined using PBS and sent to the detector (New Focus, Model 1554-B, bandwidth $\sim12$ GHz) with the same power and polarization, and the beatnote spectrum is obtained on the RF spectrum analyzer (Agilent E4405B). In the experiment, the resolution bandwidth (RBW) of the RF spectrum analyzer is 300 kHz and the video bandwidth (VBW) is 30 kHz. The scan time is 20 ms, and the frequency spanned is 20 MHz. The center frequency of the DFSSH beatnote signal is caused by an AOM frequency shift of 80 MHz, avoiding the large noise of the RF spectrum analyzer at zero and the low frequency in the DFSSH measurements. Due to the linear uncertainty of the DFSSH beatnote signal, we use the Voigt function to fit the beatnote signal. The Voigt function can be used to reflect the similarity between the spectrum and the two functions, and when the Voigt function is used for fitting, the Gaussian linewidth component and the Lorentz linewidth component can be obtained. We can obtain the laser linewidth using the following equation \cite{ref-33}:
\begin{equation}
\label{equ29}
\Delta_V=\frac{1}{2}(1.0692\Delta_L+\sqrt{0.86639\Delta_L^2+4\Delta_G^2})
\end{equation}
where $\Delta_V$ is the laser linewidth fitted using the Voigt function, $\Delta_L$ is the Lorentz component of laser linewidth, and $\Delta_G$ is the Gauss component of laser linewidth. We fit the Gauss component of the linewidth is 1.23 MHz, the Lorentz component is 0.73 MHz, and the comprehensive linewidth that can be calculated is 1.67 MHz.

\begin{figure}[H]
\centering
\includegraphics[width=13.9cm,height=5.1cm]{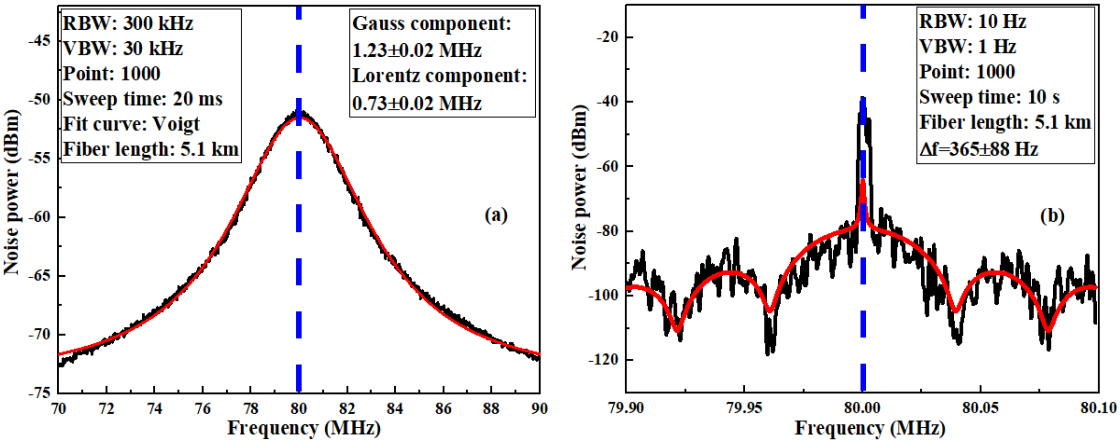}
\caption{ DBR-LD laser linewidth for the free runing case and the self-injection lock case. (\textbf{a}) DBR-LD laser linewidth for the free runing case. The RBW of the RF spectrum analyzer is 300 kHz, and the VBW is 30 kHz. The scan time is 20 ms, the frequency span is 20 MHz, and the center frequency of the DFSSH beatnote signal is caused by AOM frequency shift. The black curve is the DFSSH beatnote signal, and the red curve is the Voigt function fitting. The fitted Gauss component of laser linewidth is 1.23 $\pm$ 0.02 MHz, and the Lorentz component is 0.73 $\pm$ 0.02 MHz. Therefore, the comprehensive linewidth is 1.67 $\pm$ 0.02 MHz. (\textbf{b}) DBR-LD laser linewidth for the self-injection lock case. The injected laser is 1.3 $\upmu$W, the RBW is 10 Hz, and the VBW is 1 Hz. The scan time is 10 s, the frequency span is 200 kHz, and the center frequency of the beatnote signal is caused by AOM frequency shift. The black curve is the DFSSH beatnote signal, while the red curve is the fitting using the Equation (14), and the fitted linewidth is 365 $\pm$ 88 Hz. The blue lines in (a) and (b) are center frequency auxiliary lines.}
\label{f2}
\end{figure}

In the experiment, the high-finesse cavity we used is a mushroom-shaped flat-concave F-P cavity and made of ultra-low expansion (ULE) optical glass with a plane mirror, a curved mirror with a radius of curvature of 500 mm and length of 100 mm, and the free spectral range is 1.5 GHz. It is coated with 852 nm high-reflectivity multilayer dielectric film, and is placed in ultra-high vacuum chamber and can be precisely controlled temperature. We select suitable lens to match high-finesse F-P cavity, tuning the DBR-LD laser frequency to the cavity resonance frequency, using a lens with convex curvature radius of 300 mm to achieve the collimation of transmitted laser. There is 1.3 $\upmu$W laser is fed back into the laser diode by the PBS injection of the specially designed isolator in the self-injected lock path. We tune the driving current of DBR-LD and the path of the self-injection lock to obtain the DFSSH beatnote signal that is shown in Figure \ref{f2}b.There is a significant envelope on the beatnote signal, which proves that the fiber delay time is less than the laser coherence time and injection lock have been realized. The RBW of the RF spectrum analyzer is \mbox{10 Hz}, the VBW is 1 Hz, the scan time is 10 s, and the frequency span is 200 kHz. We use the Equation (14) to fit beatnote signal, and the linewidth obtained is 365 $\pm$ 88 Hz. It is about $1/4500$ of the linewidth for the free running case. 
\subsection{The laser linewidth versus the self-injected laser power of the DBR-LD}

We obtain the variation of DBR-LD laser linewidth with self-injection power by adjusting the $\lambda$/2 before entering the laser diode. In the experiment, the loss of optics during self-injection lock path has been considered, but due to the uncertainty of the coupling efficiency inside the laser cavity, it is difficult to evaluate the final power of the coupling laser cavity, so the injection power at this time is the power before the coupling laser cavity. As shown in Figure \ref{f3}, with the self-injected optical power increases, the laser linewidth suddenly decrease, we consider that this position is the threshold of the self-injected lock power. And then the self-injected power increases further, the laser linewidth gradually decreases. The inset shows the detailed change of laser linewidth with injected power above the injected power threshold.

\begin{figure}[H]
\centering
\includegraphics[width=7cm,height=5.22cm]{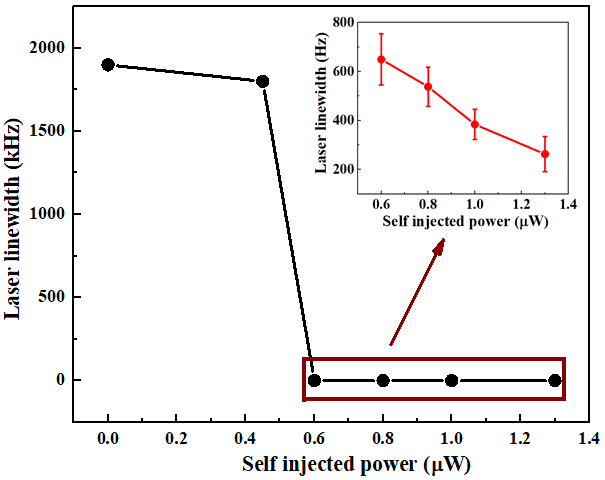}
\caption{Variation of DBR-LD laser linewidth with self-injection power. The inset shows the zooming of the laser linewidth versus self-injection laser power above the self-injection threshold.}
\label{f3}
\end{figure}
\subsection{The Self-Injection Locking Current Range of the DBR-LD versus the Self-Injected Laser Power}

The relationship between the self-injected power and the injected lock range is also tested in the experiment. Figure \ref{f4}a shows the transmission signals of the high-finesse F-P cavity (black lines) by scanning the DBR-LD driving current, the self-injected power is 1.3 $\upmu$W. It can be seen that the cavity mode can be transmitted with in certain current range, and we believe that the laser frequency in this current range is a fixed value, it is the resonance frequency of the high-finesse F-P cavity and not change with the change of the driving current. We consider this is the current range in which self-injection lock can be realized. The red line is the triangular wave signal, and the sweep current range is \mbox{20 $\upmu$A}, corresponding to the frequency range of DBR-LD in the free-running case is \mbox{170 MHz}. Figure \ref{f4}b shows the variation of  the range of injected locked currents versus different self-injected powers. It can be seen that with the increase of injection power, the injection lock range also increases. This is consistent with the theoretical analysis.

\begin{figure}[H]
\centering
\includegraphics[width=13.6cm,height=5cm]{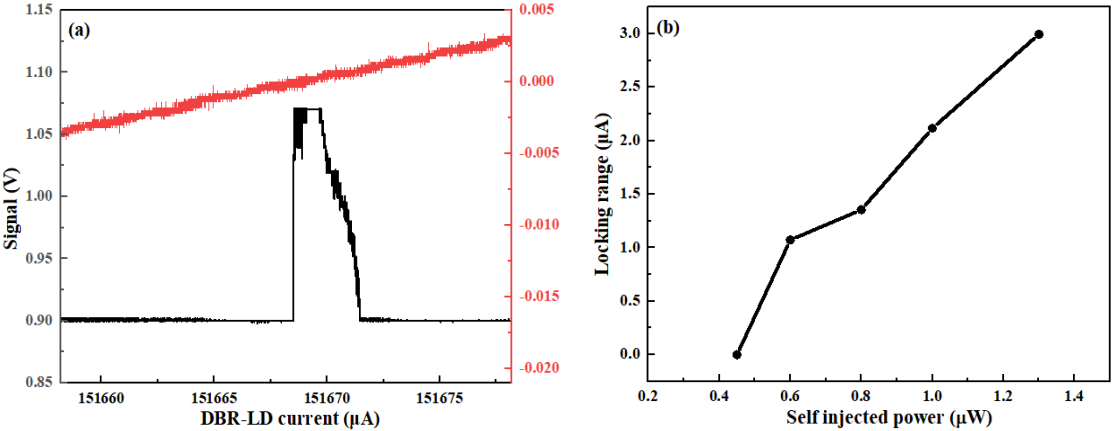}
\caption{ The high-finesse F-P cavity transmition signal for the self-injection lock case. (\textbf{a}) The high-finesse F-P cavity transmition signal (the black trace) for the injected laser power is 1.3 $\upmu$W. The red line is the triangular wave signal, and that scanning current range is 20 $\upmu$A; (\textbf{b}) Variation of the lock current range of DBR-LD with different self-injection laser power.}
\label{f4}
\end{figure}
\subsection{Measurement of the Laser Phase Noise}

To further characterize the performance of the laser diode for the self-injection lock case, we consider its phase noise variation. The transition of each atom during the process of spontaneous radiation of the carrier of the laser diode is random in space and time, and the phase and propagation direction of the optical radiation are different, so the randomness of the frequency causes the fluctuation of the phase and then forms phase noise. Linewidth is a measurement of the deviation of the optical frequency to the center caused by random fluctuations in the laser phase, so there is an intrinsic relationship between the linewidth and the phase noise and frequency noise \cite{ref-34}.

In this experiment, we use a detuned F-P cavity method to convert the phase noise into intensity noise. The free spectral range of the 852 nm F-P confocal cavity used for the phase noise measurement is 500 MHz, and the cavity length is 150 mm. The measured F-P cavity linewidth is $\sim3.5$ MHz, and the cavity finesse is 142. The intensity noise of the laser diode is close to the shot noise, and the phase noise is above its intensity noise \cite{ref-32}. Thus, we believe that the suppression mainly comes from the laser phase noise. In the zero-span mode of the RF spectrum analyzer, we scan the PZT on the F-P cavity to obtain the M-type conversion curve of the phase noise to the intensity noise for both the laser diode free running case and the self-injection lock case at different analysis frequencies. As shown in Figure \ref{f5}a, the black squares are the conversion signals of the laser phase noise to the intensity noise for the DBR-LD running free case, and the red dots are the conversion signals of the laser phase noise to the intensity noise for the DBR-LD self-injection lock case. We can see that the phase noise is well suppressed at a low frequency range. As shown in Figure \ref{f5}b, the laser phase noise is suppressed by more than 30 dB at an analysis frequency of 100 kHz . It can be seen that, although the self-injection lock technique has a poor ability to reduce the phase noise in a high frequency range, the phase noise is still lower than the DBR-LD free running case.

\begin{figure}[H]
\centering
\includegraphics[width=13.6cm,height=5.4cm]{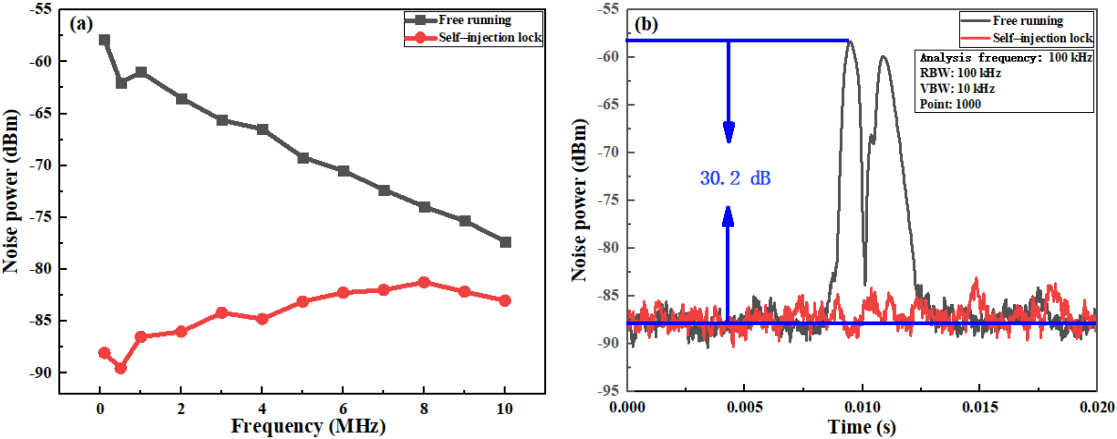}
\caption{Laser phase noise measurement. (\textbf{a}) The phase noise at different analysis frequencies is obtained by using the detuned F-P cavity method. The black squares are the phase noise for the free running case, and the red dots are the phase noise for the self-injection lock case; (\textbf{b}) The typical M-type conversion curve of the phase noise for the free running case and the self-injection lock case at the analysis frequency of 100 kHz in zero-span mode. The phase noise is clearly reduced about 30.2 dB at 100 kHz analysis frequency due to the self-injection lock (corresponding to the leftmost date in Figure \ref{f5}a). The RBW is 100 kHz and the VBW is 10 kHz.}
\label{f5}
\end{figure}
\section{Conclusions}

In conclusion, we conducted an 852 nm DBR-LD self-injection locking experiment based on the high-finesse optical F-P cavity. The DFSSH method of laser linewidth measurement was used to measure the DBR-LD laser linewidth. It was narrowed from 1.67 MHz for the free running case to 365 Hz for the self-injection lock case. Then, we also analyzed the variation in the DBR-LD laser linewidth and the injection lock current range with different self-injection lock laser power values, which was consistent with the theoretical analysis. In addition, we used a detuned F-P cavity method to convert the laser phase noise into intensity noise and compared the phase noise in the 0.1--10 MHz analysis frequency range for the laser free running case and the self-injection lock case. Especially at the 100 kHz analysis frequency, the laser phase noise was effectively suppressed by 30 dB. 

We believe that, if there are some high-finesse F-P cavities with different cavity linewidths that are used as the filtering elements and the transmitted lasers are used for self-injection locking, at the same self-injected laser power, the narrower the cavity linewidth is, the narrower the output laser linewidth will be. However, due to the optical path of the cavity filtering and the self-injection locking is in the air, it is easily affected by the external environment, introducing large phase fluctuations, which may lead to there being little difference in the output laser linewidths under cavity filtering with different linewidths. Thus, if the volume of the experimental system is reduced, the stability is improved, and an active phase feedback is performed, the final laser performance will be better. For now, we can use this narrow linewidth laser to measure the practical finesse value of the high-finesse F-P cavity, and perform two-step Rydberg excitation experiments. As a next step, this technique can also be extended to other wavelengths, such as 780 nm and 795 nm (for rubidium), 767 nm and 770 nm (for potassium), and 689 nm and 698 nm (for strontium), so that the narrow linewidth lasers obtained based on this technology can be widely used in quantum simulations, atomic clocks, and precision measurements.

\hspace*{\fill}

\noindent\textbf{Funding. }National Key Research and Development Program of China
(2021YFA1402002); National Natural Science Foundation of China
(11974226).

\hspace*{\fill}

\noindent\textbf{Disclosures. }The authors declare no conflicts of interest.

\hspace*{\fill}

\noindent\textbf{Data availability. }Data underlying the results presented in this paper may be available from the corresponding author
upon reasonable request.










\end{document}